\documentclass[preprint,5p]{elsarticle}

\expandafter\let\csname equation*\endcsname\relax
\expandafter\let\csname endequation*\endcsname\relax
\usepackage{amsmath}
\usepackage{amssymb}
\usepackage{graphicx}
\usepackage{dcolumn}
\usepackage{bm}
\usepackage[utf8]{inputenc}
\usepackage{mathtools}
\usepackage{lipsum,mathtools,cuted}
\graphicspath{{Figures/}}

\journal{Chaos, Solitons and Fractals}
\usepackage{tikz}
\usepackage{calc}

\newcommand*{\TextScale}{0.5}
\newcommand*{\SlashAngle}{45}
\newcommand*{\SlashScale}{1.3}

\newlength{\NeumeratorXShift}
\newlength{\DenomiatorXShift}
\newlength{\NeumeratorYShift}
\newlength{\DenomiatorYShift}

\tikzset{Slash/.style={scale=\SlashScale, rotate=\SlashAngle}}
\tikzset{Neumerator/.style={scale=\TextScale, xshift=-\NeumeratorXShift, yshift=\NeumeratorYShift, inner sep=0, outer sep=0}}
\tikzset{Denominator/.style={scale=\TextScale, xshift=\DenomiatorXShift, yshift=-\DenomiatorYShift, inner sep=0, outer sep=0}}

\usepackage{verbatim}

\begin{document}

\begin{frontmatter}
\title{Chaos and localization in the Discrete Nonlinear Schr\"odinger Equation}

\author[iub1,iub2]{Stefano Iubini}
\corref{corr}
\address[iub1]{Consiglio Nazionale delle Ricerche, Istituto dei Sistemi Complessi, Via Madonna del Piano 10 I-50019 Sesto Fiorentino, Italy}
\address[iub2]{Istituto Nazionale di Fisica Nucleare, Sezione di Firenze, Via G. Sansone 1 I-50019, Sesto Fiorentino, Italy}
\cortext[corr]{Corresponding author}
\author[pol]{Antonio Politi}
\address[pol]{ Institute for Complex Systems and Mathematical Biology and SUPA, University of Aberdeen, Aberdeen AB24 3UE, EU}

\date{\today}

\begin{abstract}
We analyze the  chaotic dynamics of a one-dimensional discrete nonlinear Schr\"odinger equation.
This nonintegrable model, ubiquitous in several fields of physics,
describes the behavior of an array of coupled complex oscillators with a local nonlinear potential.
We explore the Lyapunov spectrum for different values of the energy density,
finding that the maximal value of the Kolmogorov-Sinai entropy is attained at infinite temperatures.
Moreover, we revisit the dynamical freezing of relaxation to equilibrium, occurring when large
localized states (discrete breathers) are superposed to a generic finite-temperature background.
We show that the localized excitations induce a number of very small, yet not vanishing, Lyapunov exponents, 
which signal the presence of extremely long characteristic time-scales.  
We widen our analysis by computing the related Lyapunov covariant vectors, to investigate the interaction
of a single breather with the various degrees of freedom.
\end{abstract}


\begin{keyword}
Discrete Nonlinear Schr\"odinger Equation \sep Discrete breathers \sep Lyapunov spectrum\sep Lyapunov covariant vectors
\end{keyword}
\end{frontmatter}

\section{Introduction}
{
The interplay between chaotic and regular dynamics is a relevant aspect for the large majority of models in 
physics, which are neither completely integrable nor perfectly ergodic.
A famous example where a mechanism of dynamical ergodicity-breaking affects the macroscopic behavior 
is given by the Fermi-Pasta-Ulam-Tsingou model, introduced in 1954 to describe the relaxation of a chain 
of nonlinear coupled oscillators. In this case, an exceedingly slow relaxation time scale is observed as
a consequence of the quasi integrability of the dynamics at low temperatures and related to the closeness 
to the integrable Toda lattice~\cite{BCP13}. Conversely, for sufficiently
large energy densities, the system displays {\it de facto}  an ergodic behavior. 

A different scenario occurs in chains of coupled rotors, which exhibit persistent local regular orbits
in the limit of large temperatures. Such an unusual behavior is observed when a very large amount of energy is
localized on a few lattice sites, a condition that is realized by the creation of long-living discrete 
breathers states~\cite{flach98,flach08} superposed to a chaotic background. In this regime, breather states are
almost decoupled from the background and ergodization times were observed to diverge in the limit of large energy densities, while 
 the degree of chaoticity measured by the the largest Lyapunov exponent 
was found to be essentially constant~\cite{mithun19}. 

A similar phenomenon  occurs in the Discrete Nonlinear Schr\"odinger (DNLS) equation~\cite{kevrekidis09}, 
a model with several applications
in different domains of physics including nonlinear optics~\cite{christodoulides88,eisenberg98,morandotti99},
 cold atoms~\cite{smerzi97,trombettoni01,cataliotti01,cataliotti03,makarov17} and nanomagnetic systems~\cite{borlenghi14,borlenghi15}.
Here, the localization dynamics at high energy densities is driven by a condensation process: in the localized 
region, entropic arguments~\cite{RN,R1,R2} predict that  in the thermodynamic limit an extensive amount of energy 
is localized at equilibrium on a single site while the rest of the system is uniformly at infinite temperature. The nature of the 
condensation transition has been recently analyzed in the microcanonical ensemble by means of large deviations techniques, showing that
in the localized regime statistical ensembles are not equivalent and that genuine negative-temperature states
can arise for finite, still large system sizes~\cite{GILM1,GILM2}.  On top of that, purely dynamical effects were observed to
dramatically slow down the thermalization process. In~\cite{ICOPP} it was shown that the typical relaxation time-scales
increase exponentially with the breather norm even when the breather sits on a positive-temperature background, while 
in Mithun {\it et al.}~\cite{mithun18} it was given evidence that a weakly nonergodic dynamics takes place inside the localized region of the model,
undetected by the largest Lyapunov exponent.
Moreover, in Iubini {\it et al.}~\cite{CBSW} it was clarified that the slow dynamics induced by localization processes in the DNLS chain does
not only determine its  relaxation to equilibrium, but it also affects genuine steady transport regimes.

In this paper, we study the chaotic properties of a DNLS chain by both analysing the whole spectrum of Lyapunov exponents (LEs)
and the spatial structure of the corresponding covariant Lyapunov vectors (CLVs). 
As expected, there are four exactly zero LEs, associated with the conservation of the total energy and the total mass (see their definition in the following section) and representative, respectively, of the  invariance of the system under time translations and under  homogeneous  rotations in the complex plane.
Moreover, we show that the Lyapunov spectrum is very well described by a power law, identified by only
two parameters (the maximum LE and the singularity parameter) which is able to capture the spectral shape even far from the maximum.
In the high energy-density region, long times are required to let the spectral shape settle to its asymptotic form
due to the presence of long-living breathers.
As discussed in the following sections, our results are consistent with the previous numerical observations of 
genuine negative-temperature states as well as with their possible finite-size nature.

However, our main interest is the characterization of the regime emerging when a large breather is superposed to a background
at positive temperature. The mutual exchange of mass and energy between breather and background is so slow that
this regime can be considered to be (quasi)stationary and thereby compute
quantities such as the LEs which, strictly speaking, are well defined only in the infinite-time limit.
We show that the existence of quasi-conserved quantities speculated in Ref.~\cite{ICOPP} reflects itself in the presence
of additional small-amplitude Lyapunov exponents.
Conversely, the bulk of the Lyapunov spectrum involving the largest exponent (in absolute value) is essentially insensitive
to the presence of stable breather solutions, thus confirming the results obtained in~\cite{mithun18} from the analysis
of the largest exponent. 

Next, we evaluate the covariant vectors, to explore the coupling between the breather and the various stable and unstable directions
(identified by the several CLVs). The simulations confirm the intuitive idea that the coupling strength with highly unstable directions
is strongly depressed and only the weakly-unstable directions contribute to the breather dynamics.

More precisely, in Sec.~\ref{sec:MM}, we define the model, recalling some basic properties, and introduce the most important technical tools.
Sec.~\ref{sec:hom} is devoted to a discussion of homogeneous chains both in the region of positive and negative temperatures.
In Sec.~\ref{sec:clv} we discuss a setup where a breather is sitting in the middle of the chain, determining both the Lyapunov spectrum and analysing 
the CLVs.
Finally, in Sec.~\ref{sec:concl} we summarize the implications of the various simulations and recall the major open problems.

\section{Model and methods}
\label{sec:MM}

We consider the one-dimensional DNLS equation with fixed boundary conditions
\begin{equation}
\label{eq:dnls}
i\dot{z}_j = -2 |z_j|^2 z_j -z_{j-1} -z_{j+1}\,,
\end{equation}	 
where  $j=0,\cdots,N-1$ and $z_j$ are complex-valued amplitudes (with $z_{-1}=z_N=0$).	 
Upon recognizing that $(z_j,iz_j^*)$ are a proper set of canonical variables, Eq.~(\ref{eq:dnls}) can be derived 
from the Hamiltonian
\begin{equation}
\label{eq:H}
\mathcal{H}= \sum_{j=0}^{N-1} |z_j|^4 + (z_j^* z_{j+1} + z_j z_{j+1}^*)
\end{equation}
through the Hamilton equations $\dot{z}_j=-\partial \mathcal{H} / \partial (iz_j^*)$. Besides the total energy $\mathcal{H}$, the model possesses 
a second exactly conserved quantity, the total mass (norm)
\begin{equation}
\mathcal{A}=\sum_{j=0}^{N-1} |z_j|^2\,.
\end{equation}	

Thermodynamical equilibrium states of the model can be described in terms of the energy density $h=\mathcal H/N$ and 
the mass density $a=\mathcal A/N$, and they can be represented as single points in the plane $(a,h)$.
A derivation of the equilibrium phase diagram in the $(a,h)$ space was done in Rasmussen {\it et al.}~\cite{RCKG00} within
the grand-canonical ensemble and it is shown in Fig.~\ref{fig:phase_diag}
 \begin{figure}[ht]
\includegraphics[width=0.45\textwidth,clip]{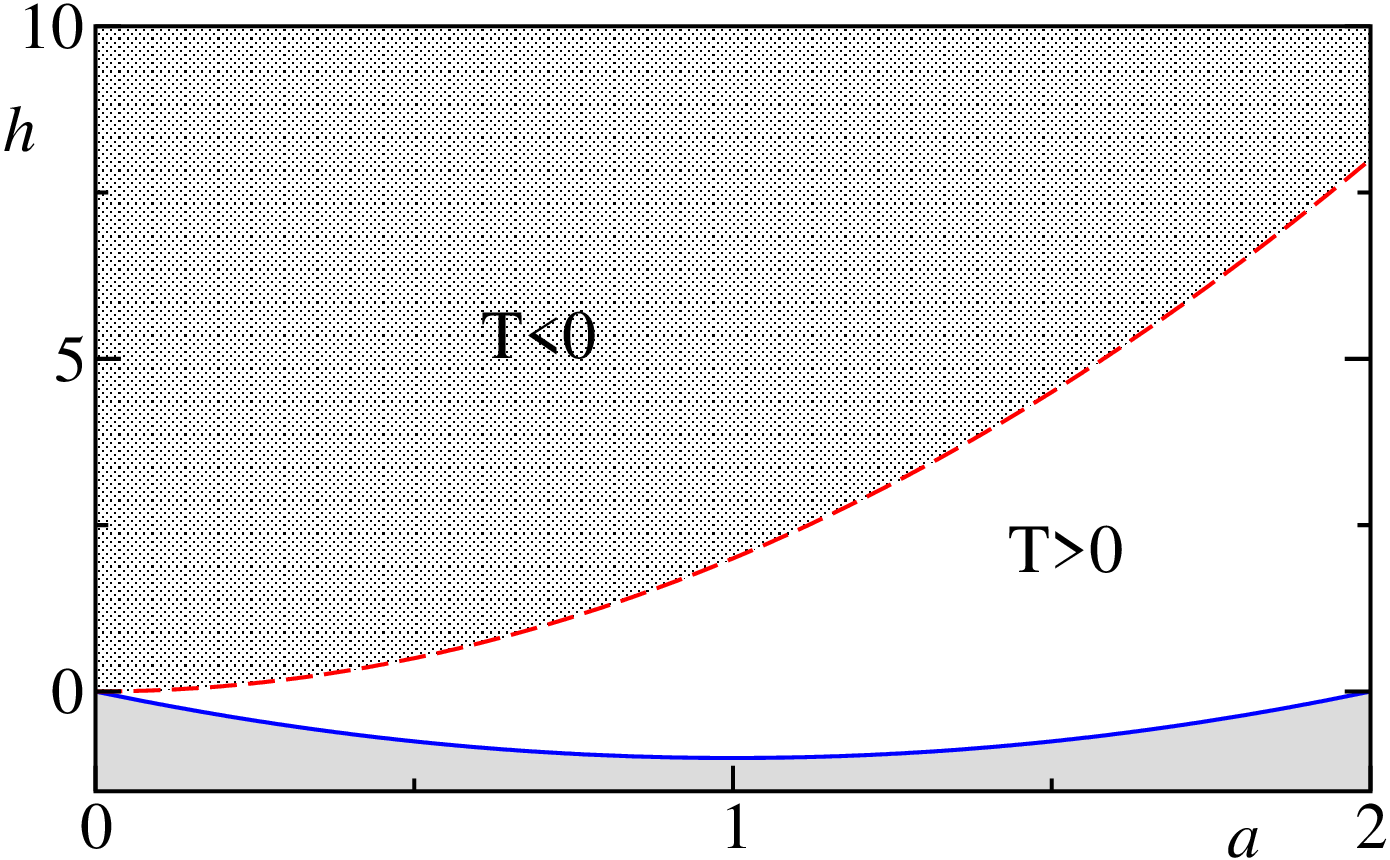} 
 \caption{Schematic phase diagram of the DNLS equation in the $(a,h)$ plane as obtained in Rasmussen {\it et al.}~\cite{RCKG00}. The blue solid line identifies the set of the zero-temperature ground states. The red dashed curve
corresponds to infinite temperature.}
 \label{fig:phase_diag}
 \end{figure}
The lower solid line corresponds to the ground state $(T=0)$  and it corresponds to the relation $h=a^2-2a$.
Below this curve, no accessible states are present. Positive temperature states are located in the region
between the $T=0$ line and the infinite-temperature line (red dashed), defined by $h=2a^2$. Finally, the states above the infinite
temperature line belong to the so-called negative-temperature region and display energy localization~\cite{IFLOP}, nonequivalence of ensembles~\cite{GILM1,GILM2} and weak ergodicity breaking~\cite{mithun18}.

As anticipated, we use LEs~\cite{Piko-Politi}
to characterize the different regimes and more specifically, to investigate the weak coupling between
single tall breathers and the surrounding background.

The phase space of a DNLS chain of size $N$ is $2N$-dimensional. Hence, there exist $2N$ LEs	 $\lambda_m$ that
are customarily ordered from the largest one $\lambda_1$ down to the most negative one $\lambda_{2N}$. 
Hamiltonian systems such as the DNLS equation are characterized by symmetric spectra, i.e. $\lambda_m = -\lambda_{N-m+1}$. 
Moreover, because of two symmetries
(invariance under time translation, invariance under a homogeneous rotation of the $z_j$s in the complex plane), 
 four exponents 
are expected to vanish.
In the large $N$ limit we expect the LEs to depend on $m$ and $N$ via the compound variable
$\rho = (m-1/2)/N$: $\lambda_m = \Lambda(\rho= (m-1/2)/N)$, where the $1/2$ shift is a correction term introduced to highlight the perfect symmetry $\Lambda(\rho) = \Lambda(2-\rho)$ for finite $N$ values.
This scaling behavior has a simple physical interpretation:
the degree of chaos exhibited by a system of size $N$ is proportional to $N$.
For instance, an upper bound to the Kolmogorov-Sinai entropy, quantified by the Pesin relation~\cite{eckmannREV},
 can be written as
\begin{equation}
K_{K\!S} = \sum_{m=1}^N \lambda_m = k_{K\!S} N\,,
\end{equation}
where
\begin{equation}
k_{K\!S} = \int_0^1 d\rho \Lambda(\rho) 
\end{equation}
can be interpreted as the Kolmogorov-Sinai entropy density. We employ $k_{K\!S}$ as a global indicator of the 
chaotic properties of the DNLS model.

Finally, we study also CLVs. They identify the orientation of the different expanding and
contracting directions in the phase space. A detailed description of the various algorithms proposed to determine CLVs can be found in Pikovsky and Politi~\cite{Piko-Politi}.
Previous studies of spatially extended Hamiltonian systems and more in general of space-time chaos, have revealed that CLVs are
typically localized in physical space and that the center of localization fluctuates in time 
(see for instance \cite{Pi-Po98})
This reflects the continually changing location of the centers of instability in homogeneous systems.
In the present context, a breather breaks the homogeneity and it is therefore interesting to investigate the implications
of this phenomenon.

\section{The homogeneous case}
\label{sec:hom}

We start our numerical analysis by computing  Lyapunov spectra for two energy densities and a fixed
mass density $a$, set equal to 1 (this is the value selected in several previous simulations of DNLS chains).
The spectra reported in Fig.~\ref{fig:lyap_gen}(a) have been obtained for $h=1.65$, a value which corresponds\footnote{The 
temperature corresponding to the value $h=1.65$ has been measured from equilibrium simulations by employing a suitable microcanonical definition of temperature, see~\cite{iubini12} and references therein for details. More recent developments on the mapping of thermodynamic parameters of the DNLS model are also in Levy and Silberberg~\cite{levy18}.}
 to the positive temperature $T\simeq 13$.
We plot only the first part of the spectrum, the second part being equal with opposite sign.
The three sets of symbols, namely circles, diamonds and triangles, refer to $N=49$, 98, and 196, respectively. The good overlap confirms the validity of the expected
scaling behavior. The last two zero exponents are hardly recognizable as such, but they are there (this issue will be more
thoroughly explored in the next section).
 \begin{figure}[ht]
\includegraphics[width=0.45\textwidth,clip]{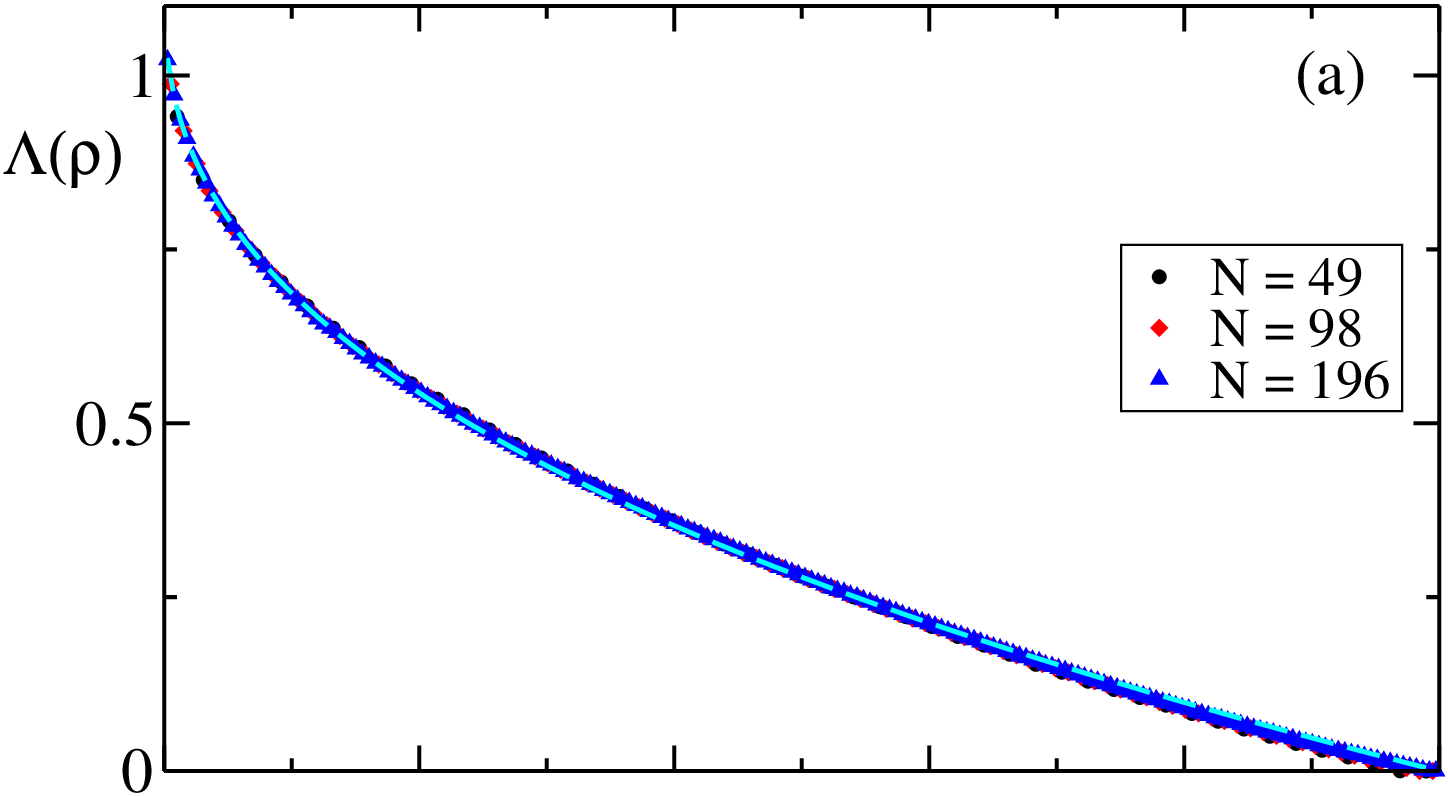}
\includegraphics[width=0.45\textwidth,clip]{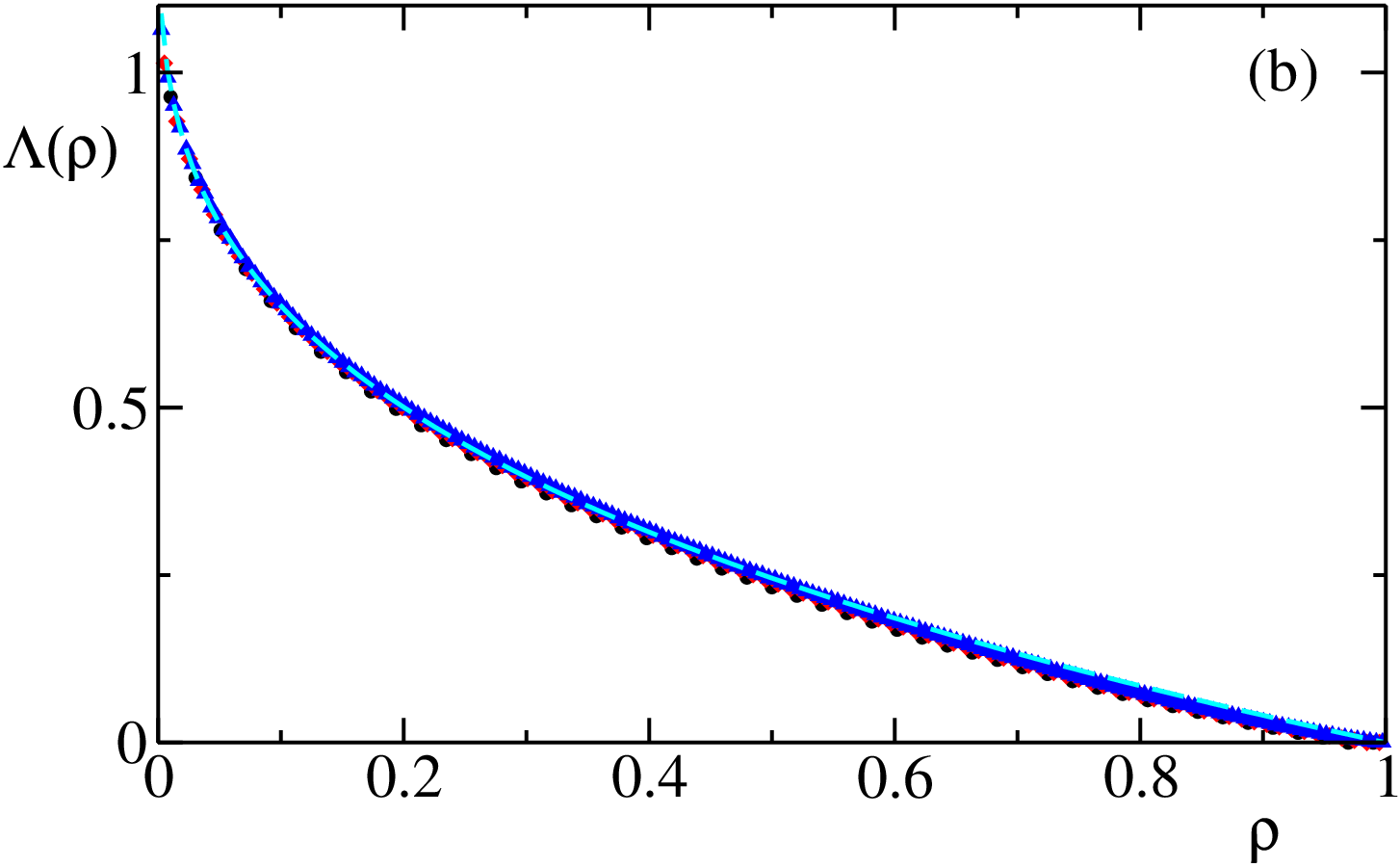}
 \caption{ Rescaled Lyapunov spectra $\Lambda(\rho)$ for energy density $h=1.65$ (panel a) and  $h=2.6$ (panel b)
  and different   system sizes. Cyan dashed lines correspond to fits with the function in Eq.~(\ref{eq:fit})
   specified by parameters $(\Lambda_0=1.12,\alpha=0.41)$ for panel (a) and $(\Lambda_0=1.3,\alpha=0.3)$ for panel (b).
 } 
 \label{fig:lyap_gen}
 \end{figure}
Remarkably, the whole spectrum is well fitted by the two-parameters function
\begin{equation}
\Lambda(\rho) = \Lambda_0(1-\rho^\alpha)\,,
\label{eq:fit}
\end{equation}
where $\Lambda_0$ is the maximum exponent, while $\alpha$ represents the degree of singularity of the spectrum in the vicinity of the maximum.
The same spectral shape is observed above the critical energy density ($h=2$) in the negative-temperature region. The results for $h=2.6$ are 
plotted in Fig.~\ref{fig:lyap_gen}(b). The maximum exponent is slightly larger than before and the singularity is more pronounced:
the effective exponent $\alpha$ is very close to 1/3, to be compared with the previous value close to 0.4.

In order to illustrate the overall energy dependence of the linear stability, we focus on a single global indicator: the Kolmogorov-Sinai entropy density $k_{K\!S}$.
 \begin{figure}[ht]
\includegraphics[width=0.45\textwidth,clip]{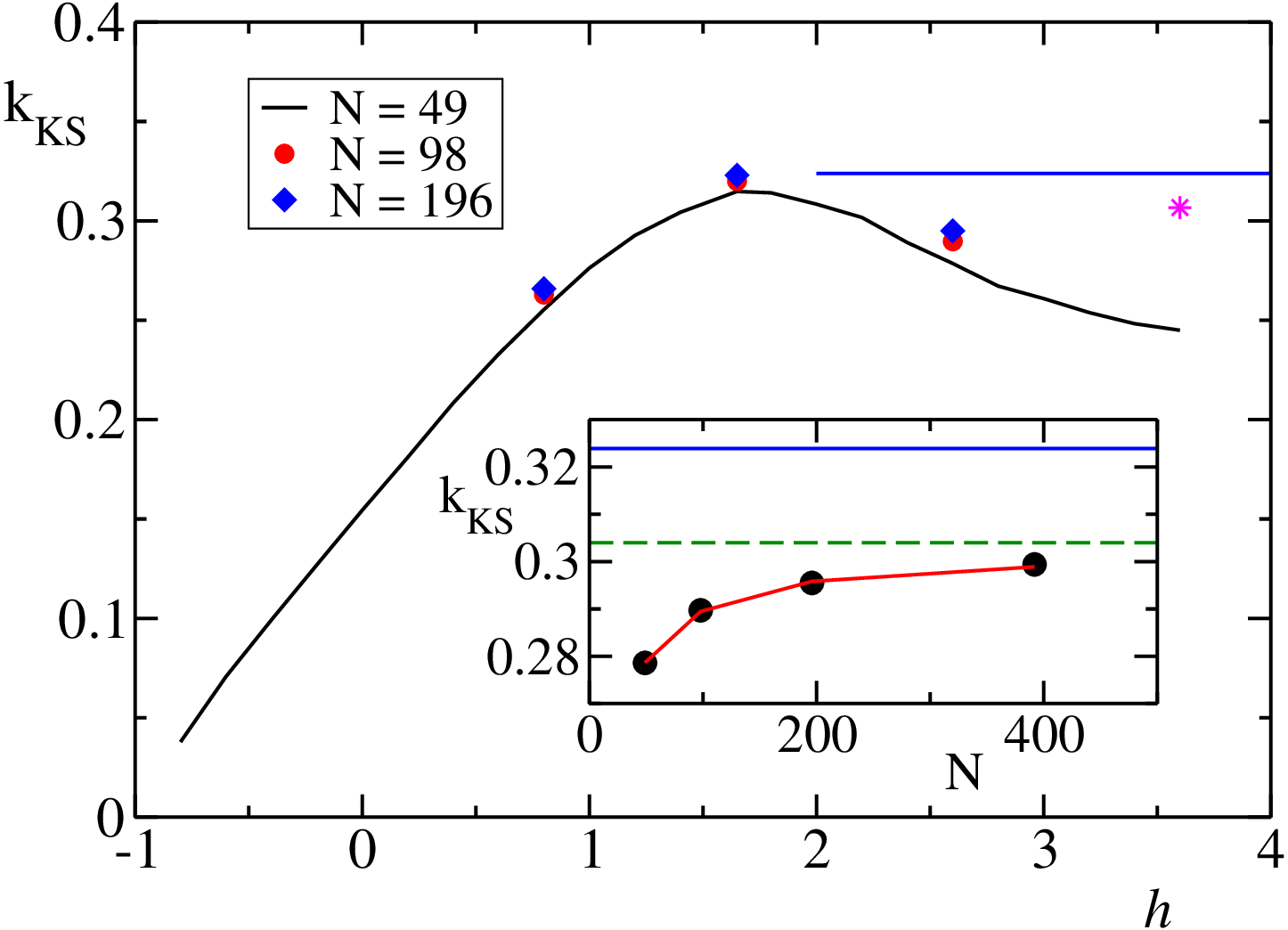}
 \caption{Kolmogorov-Sinai entropy density versus energy density for different sizes. The horizontal line shows the estimate of the maximum value $k_{K\!S}^\infty$.
 The asterisk corresponds to the value of $k_{K\!S}$ computed  at $h=3.6$ for an initial condition displaying
 a single breather superposed on an infinite-temperature background.
 Inset: $k_{K\!S}$ versus $N$ for $h=2.6$ (black circles) and fit with a law $h_{K\!S}=c_0-c_1/N^\beta$, with parameters
 $c_0=0.305$,  $c_1=0.59$ and $\beta=0.81$ (red solid line). The horizontal dashed  and solid lines are respectively the fitted asymptotic value $h_{K\!S}=a$ and the maximum entropy density $k_{K\!S}^\infty$.
 }
 \label{fig:KS}
 \end{figure}
The solid line in Fig.~\ref{fig:KS} has been obtained from simulations with $N=49$. The qualitative behavior is understandable. On the one hand,
$k_{K\!S}$ vanishes when the energy density approaches $-1$, the value corresponding to a zero-temperature
state (see Fig.\ref{fig:phase_diag}), 
where the  dynamics is regular (harmonic). On the other hand, the decrease observed at high energy-density is the consequence 
of the appearance of breathers which tend to regularize the dynamics.

Simulations performed with longer chains (red circles and blue diamonds correspond to $N=98$ and $N=196$, respectively) for $h=0.8$, 1.65 and 2.6
show that the overall shape of $k_{K\!S}$ is almost asymptotic.
The maximum of the entropy, observed around $h=1.7$ is probably a 
finite-size effect: simulations performed with longer chains (up to $N=392$) suggest
that the maximum progressively shifts towards $h=2$, i.e. at infinite temperature. 
Accordingly, we will refer to this maximum as $k_{K\!S}^\infty$.

The negative-temperature region requires a specific analysis.
As recalled in the introduction, Rumpf suggested that above
$h=2$, the asymptotic dynamics consists of a single breather surrounded by a background at infinite temperature. 
On the other hand, numerical studies performed in~\cite{IFLOP,mithun18} give evidence of a nontrivial weakly-ergodic 
where breathers are born and die (at least slightly above $h=2$).
Finally, recent more advanced theoretical arguments indicate that this regime might be a finite-size effect~\cite{GILM1,GILM2}.

It is therefore worth exploring the problem with the help of Lyapunov spectra.
A single breather cannot affect an extensive observable such as the Kolmogorov-Sinai entropy. 
Therefore, we would expect the Rumpf argument to imply an entropy density  constant above $h=2$ and equal to the infinite-temperature 
case.
In order to test the meaningfulness of this prediction, we have performed a simulation for $N=392$, starting 
from an initial condition composed of a background at infinite temperature and a single breather tall enough to ensure $h=3.6$.
The resulting $k_{K\!S}$value is denoted by the asterisk in Fig.~\ref{fig:KS}: it is slightly below our best estimates of $k_{K\!S}^\infty$.
The deviation is probably a finite-size effect.

Hence the question arises whether or not the deviation between the black solid curve and the horizontal line is a finite-size correction.
One must be careful since two types of finiteness play a role: finite-length and finite-time.
In particular, as from~\cite{ICOPP}, we know that the taller the breathers, the less they are coupled with the background.
Therefore, we have decided to focus on $h=2.6$, a not-too-large energy density where a nontrivial chaotic regime
was found in~\cite{mithun18}.
For not too large $N$ we can be confident that the time scale accessible for our simulations (a few $10^5$ units)
is sufficiently long
to be asymptotic. The resulting $k_{K\!S}$ values for different sizes $N$  are reported in the inset of Fig.~\ref{fig:KS}. There
we also fit $k_{K\!S}$ with a law $c_0-c_1/N^{\beta}$ to estimate the asymptotic  value $c_0$; it turns out to be 0.305 (see the
horizontal dashed line), significantly lower than the upper horizontal line, which corresponds to our best estimate
of the infinite-temperature entropy.

Altogether, our simulations confirm the existence of a non standard chaotic regime.

\section{Localized states}
\label{sec:clv}
In this section we discuss the stability of a DNLS chain in the presence of a single breather surrounded by a chaotic background.
Even though one expects the breather to
progressively lose mass and energy towards the background, the process is so slow~\cite{ICOPP} that it makes sense to consider the regime as stationary and thereby
compute the corresponding Lyapunov spectrum.

We have studied a chain with $N=49$ sites and energy density $h=1.65$. The resulting spectrum is plotted in Fig.~\ref{fig:brt} (solid line);
afterwards, we have added a breather of either mass 20 (circles), or 30 (diamonds) in the middle of the chain.
\begin{figure}[ht]
\includegraphics[width=0.48\textwidth,clip]{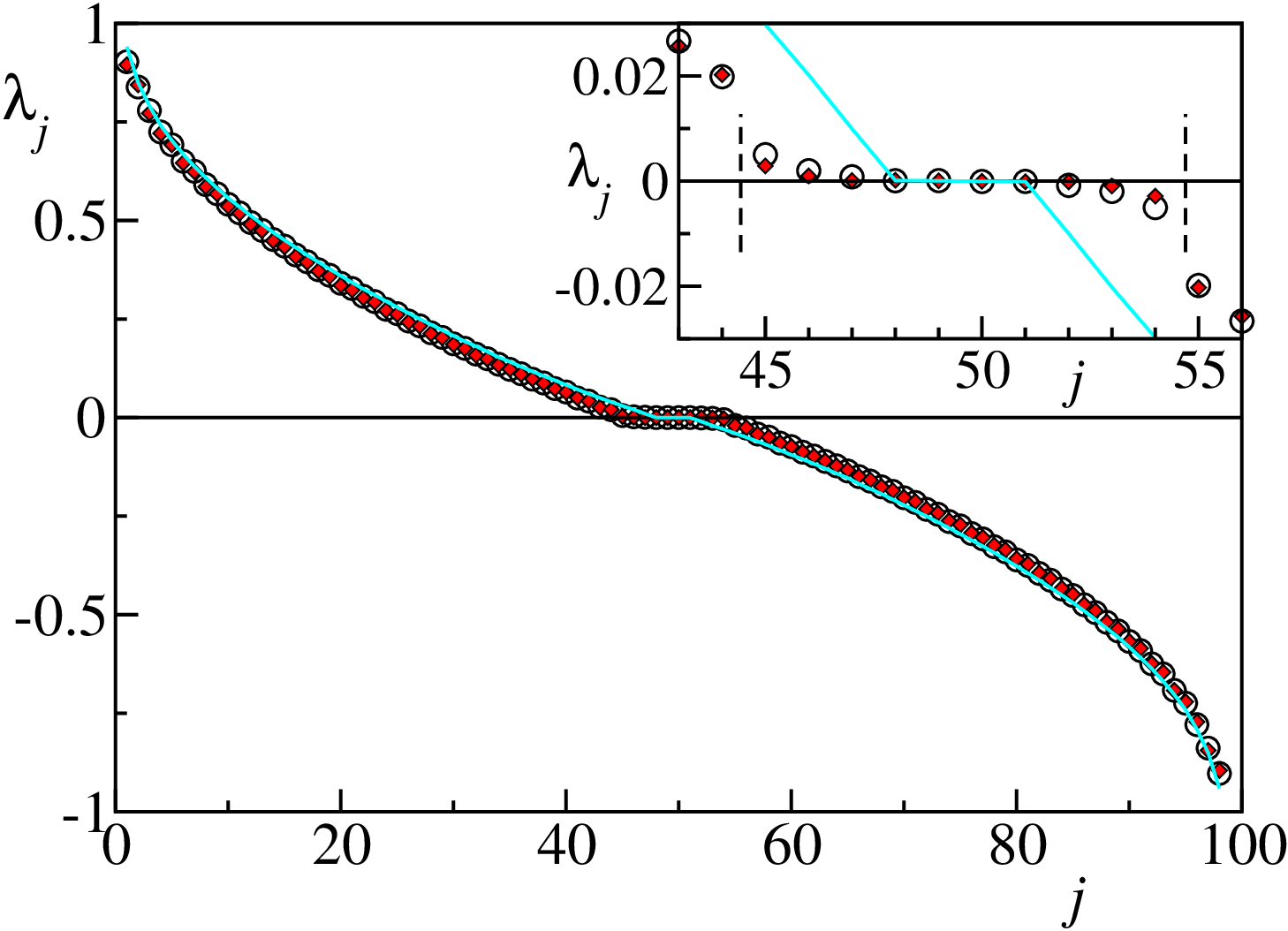} 
 \caption{Lyapunov spectrum in the presence of a single breather in a chain with $N=49$ lattice sites. Solid cyan line identifies the reference
 state with $h=1.65$ and no added breather. Symbols  correspond to the dynamics after that the reference  state is modified by the superposition on the central lattice site of a breather  of mass 20 (black circles) and 30 (red diamonds). Inset: zoom of the central region of the spectra.  When the breather is present, the region inside the two vertical dashed lines contains 10 almost vanishing LEs. }		
 \label{fig:brt}
\end{figure}
All the spectra nearly superpose to one another, confirming the expectation that a
localized ``defect" (the breather) does not modify the general structure of the spectrum.
Anyhow, the breather breaks the chain into two almost uncoupled subchains. This means that each subchain should
exhibit four (almost) vanishing exponents. Additionally, the breather itself behaves almost periodically and, being described by
two variables, we expect two further nearly zero exponents. Altogether, from this simple argument
one expects to find  ten nearly vanishing exponent.
This is indeed confirmed by looking at the inset, where we recognize ten almost vanishing exponents for both breather amplitudes.
In contrast, the spectrum in the absence of breathers (solid line in the inset) contains only four
(exact) zero exponents.

Once clarified that  the Lyapunov spectra reveal the emergence of six additional slow degrees of freedom,
a natural question concerns their spatial properties.  
 The best way to extract useful information is by determining the associated covariant
Lyapunov vectors (CLVs). 

Let $u_j(m)$ denote the (complex) amplitude of the $m$th CLV and assume that all CLVs are normalized
\begin{equation}
\sum_{j=1}^N |u_j|^2(m) =1\,,
\end{equation}
so that we can interpret $|u_j|^2$ as a probability and compute the normalized entropy
\begin{equation}
\eta(m) = - \frac{\sum_{j=1}^N |u_j|^2(m) \ln |u_j(m)|^2}{\ln N}\,.
\end{equation}
By construction, $\eta \in [0,1]$: 0 corresponds to a perfectly localized distribution (i.e. only one component different 
from zero), while
$\eta=1$ corresponds to a perfectly homogeneous distribution.
As the CLVs display different orientations depending on the current phase-space configuration, it
is useful to consider $\langle \eta \rangle$, where the angular brackets denote a time average.

 \begin{figure}[ht]
\includegraphics[width=0.48\textwidth,clip]{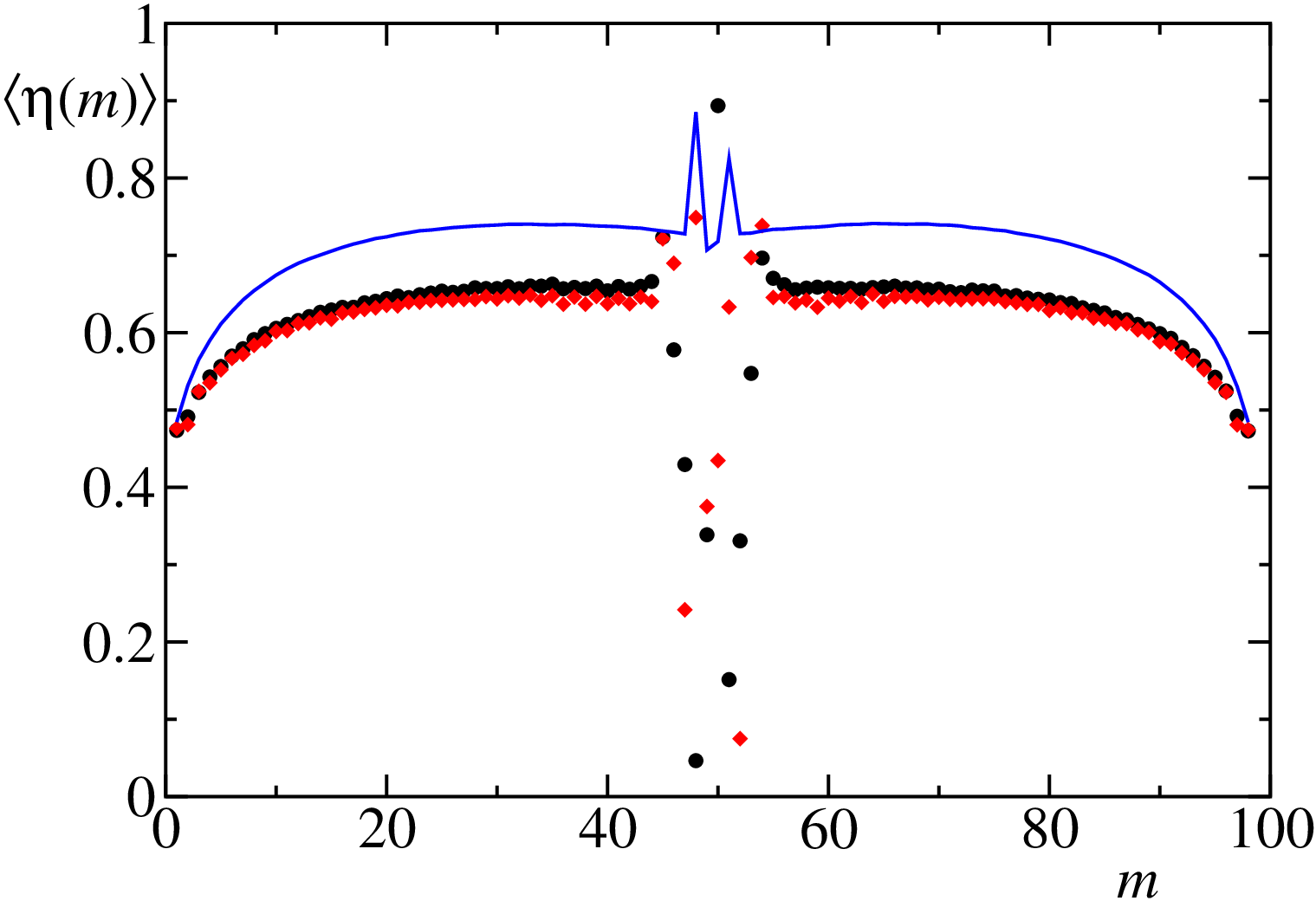} 
 \caption{Entropy $\langle\eta(m)\rangle$ of covariant vectors $u_j(m)$ for the same setup of Fig.~\ref{fig:brt}.
  Solid line, black circles and  red diamonds refer respectively to the unperturbed dynamics (no breather),
  breather mass 20, breather mass 30.}
 \label{fig:entropy}
 \end{figure}

In Fig.~\ref{fig:entropy}, $\langle \eta \rangle$ is plotted for all CLVs.
The solid curve refers to a fully homogeneous case with $h=1.65$. 
With the exception of the central four CLVs, the spectrum is perfectly symmetric: this follows from the  
symplectic structure of the underlying dynamics. 
The entropy of the central four points is not fully reliable since the corresponding
Lyapunov exponents are all equal to one another and equal to 0. In the presence
of a degeneracy, the algorithm for the computation of the CLVs is able to identify  only the space
spanned by the various vectors, but it cannot distinguish form one another. 
Anyhow, we can at least conclude  that such vectors are more extended than all the others, as
expected for directions associated to conservation laws~\cite{Piko-Politi}.
Black circles and red diamonds refer to the setup with a breather of mass 20 and 30, respectively.
In this case, the presence of ten nearly zero exponents
leads to an equal amount of fluctuating (ill-identified) entropies, easily recognizable in the middle of the
spectrum. A few of them are very localized, presumably around the breather itself.
Additionally, the breather lowers the entropy of CLVs corresponding to ``fast'' directions (i.e. 
to nonvanishing LEs): 
it essentially means that the ``fast'' CLVs become more localized.
We argue that the reason of such localization is lattice ``splitting" into two 
weakly interacting subsystems. This conjecture is confirmed by the comparison of two snapshots of 
the 24th CLV taken at different times (see Fig.~\ref{fig:vec24}): in one case the vector is localized in the
first half, while in the other it is localized in the second half.

\begin{figure}[ht]
\includegraphics[width=0.48\textwidth,clip]{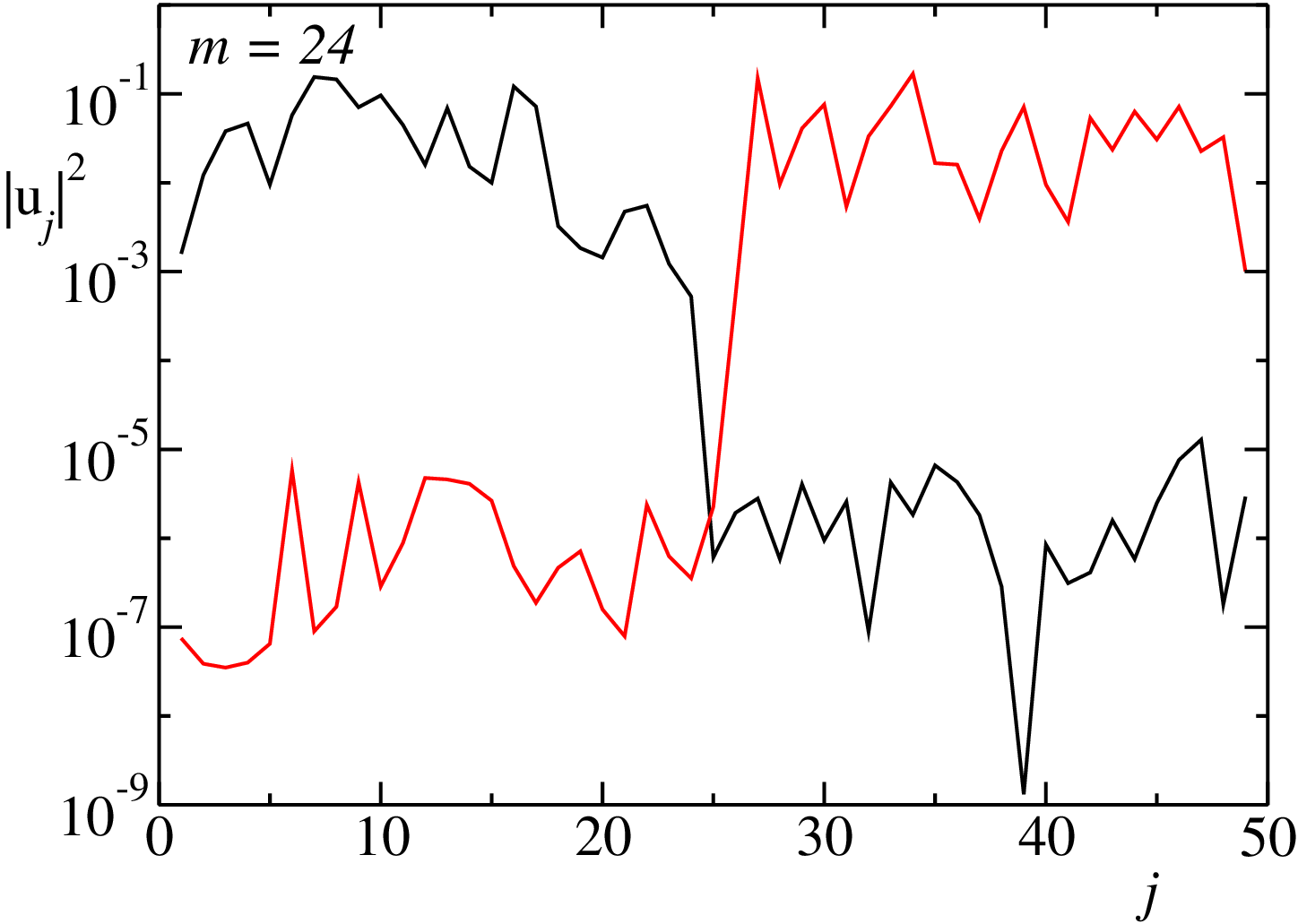} 
 \caption{Spatial profiles of the 24th CLV  
 in a chain with $N=49$ lattice sites and  in the presence of a breather with mass equal to 20 in the middle of the chain.
 The reference background configuration corresponds to $h=1.65$. The two lines correspond to two independent configurations
 of the CLV sampled during the quasi-stationary dynamics.
 }
 \label{fig:vec24}
 \end{figure} 
  
In order to clarify the role of the breather in the CLV structure, we have computed yet another
indicator, the average projection of each CLV on the position $b$ of the breather
\begin{equation}
P(m) = \langle \ln |u_b(m)|^2 \rangle
\end{equation}
where the angular brackets denote again a time average.

 \begin{figure}[ht]
\includegraphics[width=0.48\textwidth,clip]{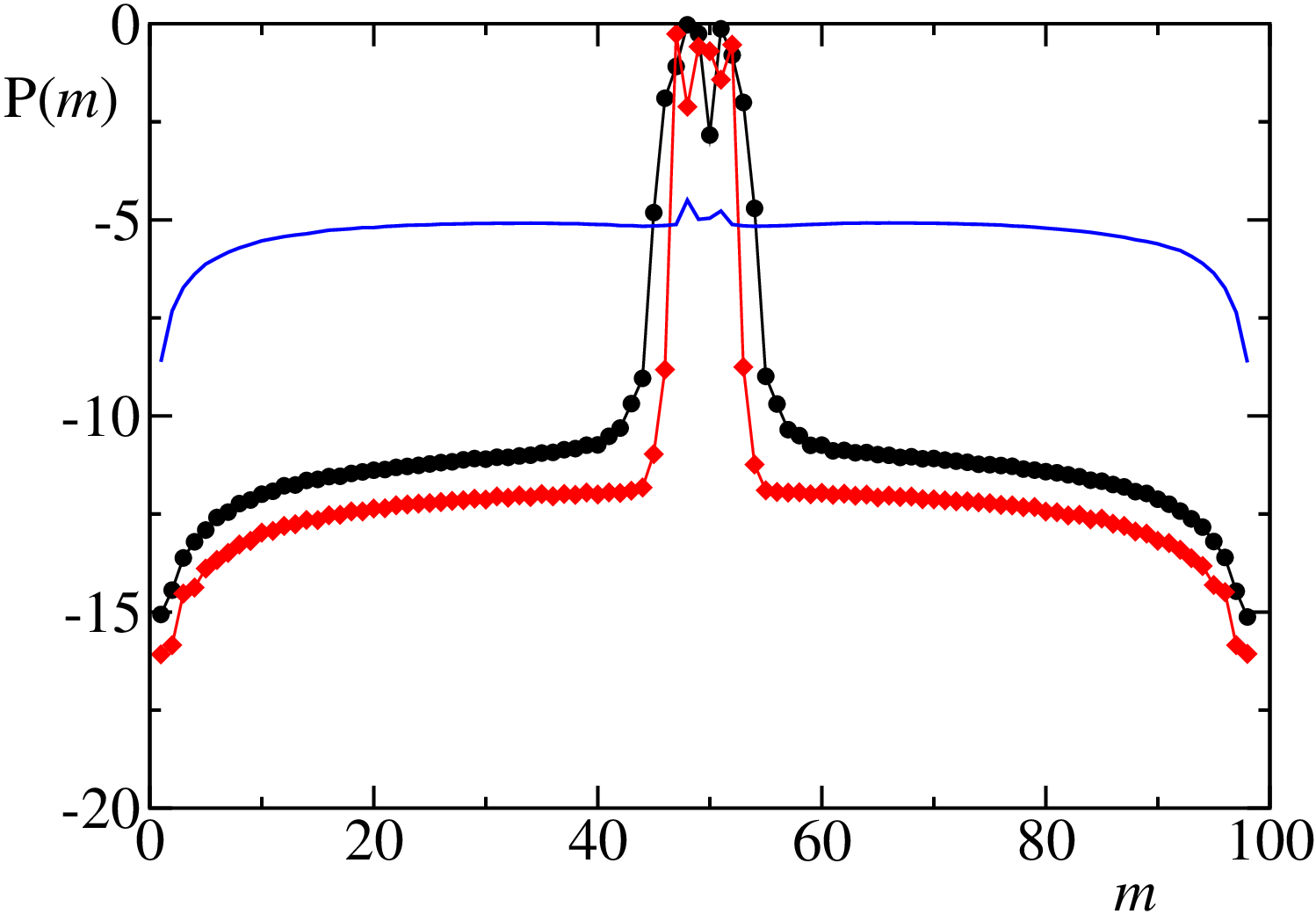} 
 \caption{Projection on the breather site $P(m)$ of CLVs for the same setup of Fig.~\ref{fig:brt}.
  Solid line, black circles and  red diamonds refer respectively to the unperturbed dynamics (no breather),
  breather mass 20, breather mass 30.
 }
 \label{fig:covar}
 \end{figure}
The results are plotted in Fig.~\ref{fig:covar}. The solid blue curve corresponds to a homogeneous chain
and reveals that almost all 
 CLVs have the same average amplitude on the breather site, with a projection 
 close to $\ln (1/N)$. Significant deviations from this value are observed only for directions corresponding to
 LEs near the boundaries of the Lyapunov spectrum.
The picture is dramatically different in the presence of the breather.  
Black circles and red diamonds, referring respectively to a breather with mass 20 and 30, display a much lower
plateau, thus implying that most of the CLVs have a negligible component on the breather site.
On the other hand, the sharp central peak shows that
the vectors corresponding to nearly zero LEs (including exactly zero LEs) are heavily localized
on the breather site.
The width of the peak is again equal to 10 and confirms that this localization property pertains exclusively 
to the emerging slow degrees of freedom and exact conservation laws.

\section{Discussion and open problems}
\label{sec:concl}

 We have extensively investigated the chaotic properties of a one-dimensional DNLS equation~(\ref{eq:dnls}) by computing the
 whole spectrum of Lyapunov exponents and the corresponding set of covariant Lyapunov vectors in various
 regimes. This analysis was motivated by the peculiar properties of the DNLS model, which exhibits 
 a condensation transition for $h>2a^2$~\cite{GILM1,GILM2}. In the so-called localized phase, it is 
known that local regular excitations (discrete breathers) spontaneously emerge  out of a uniform incoherent background, with
 signatures of non ergodicity~\cite{mithun18,arezzo21} and extremely slow relaxation timescales~\cite{ICOPP}.    
 
 Our study focused on three different regimes: 
{\it i}) uniform  states in the delocalized region $h<2a^2$; 
{\it ii})  partially delocalized states slightly above the critical line $h=2a^2$; 
{\it iii}) states composed of a very tall breather superposed to a delocalized background. 
 In all these regimes, the Lyapunov spectra show the clear presence of {\it extensive chaos}.
 The Lyapunov spectrum is fitted remarkably well by the two-parameter function  $\Lambda(\rho) = \Lambda_0(1-\rho^\alpha)$,
where $\Lambda_0$ is the maximum Lyapunov exponent, while the singularity exponent $\alpha$ encodes the shape of the
spectrum.
It would be interesting to explore whether this representation extends to a mass density different from 1.

For homogeneous regimes, the Kolmogorov-Sinai entropy density $k_{K\!S}$  
displays a non monotonic behavior as a function of the energy density $h$
(and fixed $a=1$). It increases up to $h=2$ (i.e. for $T \to \infty$), thereby
exhibiting a smooth but consistent decrease.
This is in contrast with some theoretical expectations:
entropic considerations indeed suggest that the ``excess'' energy should condensate
on a single breather~\cite{RN,R1,R2} surrounded by an infinite-temperature background.
Accordingly, $k_{K\!S}(h)$ should be independent of $h$ and equal to $k_{K\!S}(2)$.
In fact, as also confirmed by our simulations, the presence of a breather
cannot modify an intensive observable such as $k_{K\!S}$ (for large enough system sizes).

The decrease of $k_{K\!S}(h)$ can be understood by invoking
the finite-size scaling analysis carried out in Ref.~\cite{GILM1},
where, with the help of large-deviations techniques, it was shown that an extended negative-temperature 
region is present up to an energy density $h_c \approx h_c = 2 + \zeta_c N^{-1/3}$, 
with $\zeta_c\simeq 11$. According to this formula, which is
valid close to the infinite-temperature line~\cite{GILM1},    
the single-site localization arises for $N \gtrsim 6000$ for $h=2.6$ and for $N\gtrsim 300$ for $h=3.6$, both above
the sizes considered in our study. 
In this light, our results, based on the computation of the Kolmogorov-Sinai entropy, provide 
an independent, purely dynamical indication of the existence of stable, extended negative-temperature states
in the DNLS equation.
Finally, one should not neglect the presence of exponentially long time-scales which may hinder the observation
of a truly asymptotic behavior. Such difficulties are expected to arise for yet larger systems sizes than
those considered in this paper.

Although the presence of a tall breather does not affect the Kolmogorov-Sinai entropy density, it has important dynamical implications.
Besides the four exactly vanishing exponents induced by conservation laws, we have found six additional 
nearly vanishing exponents, which signal the presence of very long time-scales: this is because the breather
effectively breaks the DNLS chain into two almost decoupled subchains.

The impact of the breather over finite time scales is better captured by the CLV analysis.
The covariant vectors associated to the small LEs are strongly localized on the breather site, 
while the projection of the ``fast'' CLVs on the breather site is several orders of magnitude smaller.
Moreover,  the overall degree of localization of fast CLVs, here quantified through the normalized 
entropy parameter $\eta$, was found to increase due to a phenomenon of confinement: essentially the breather forces 
fast CLVs to stay in either one of the two weakly-interacting subchains, 
although ``jumps" from one subchain to the other one can sporadically occur.
 
How is this Lyapunov analysis related to previous observations of frozen relaxation dynamics 
induced by breather excitations?
In Ref.~\cite{ICOPP} it was conjectured that the very weak relaxation of the breather amplitude
might be related to the existence of an additional quasi-conserved quantity (a so-called
adiabatic invariant).
Our Lyapunov analysis has shown the presence of six slow degrees of freedom, which are however
interlaced with the conserved quantities.
A more quantitative analysis is hindered by the very nature of the Lyapunov exponents. 
We have computed the LEs under the assumption of an underlying (pseudo)stationary regime, which is
certainly valid over the time-scales considered in this paper. Unfortunately, the resulting statistical
fluctuations are too large to estimate the order of magnitude of the small LEs.
On the other hand, longer simulations would break the stationarity hypothesis. Hopefully one might perform ensemble averages:
this possibility should be explored in future studies.

Additional, enlightening information can come from the CLVs and more precisely from the negligible
effective interaction of the ``fast" CLVs with the breather site, testified by the exponentially small projection
on the breather site. Novel ideas are however required to transform this observation into a quantitative argument.

\bibliography{biblio}

\end{document}